\documentclass[12pt]{article}
\def\theequation{\arabic{section}.\arabic{equation}}
\usepackage{amssymb}

\newcounter{rown}

\begin{document}
\renewcommand{\thefootnote}{\fnsymbol{footnote}}
\renewcommand{\theequation}{\thesection.\arabic{equation}}
\title{On Supergroups with Odd Clifford Parameters and Supersymmetry with Modified Leibniz Rule.}
\author{Z. Kuznetsova${}^{a}$\thanks{{\em e-mail: zhanna@cbpf.br}}~, M. Rojas${}^{b}$\thanks{{\em e-mail: 
mrojas@cbpf.br}} 
and F. Toppan${}^{b}$\thanks{{\em e-mail: toppan@cbpf.br}}
\\ \\
${}^a${\it ICE-UFJF, cep 36036-330, Juiz de Fora (MG), Brazil}
\\ 
${}^b${\it CBPF, Rua Dr.}
{\it Xavier Sigaud 150,}\\ {\it cep 22290-180, Rio de Janeiro (RJ), Brazil}}
\maketitle
\begin{abstract}
We investigate supergroups with Grassmann parameters replaced by odd Clifford parameters.
The connection with non-anticommutative supersymmetry is discussed. A Berezin-like calculus
for odd Clifford variables is introduced. Fermionic covariant derivatives for supergroups with odd Clifford variables are derived. Applications to supersymmetric quantum mechanics are made.
Deformations of the original supersymmetric theories are encountered when the fermionic covariant derivatives
do not obey the graded Leibniz property. The simplest non-trivial example is given by the $N=2$ SQM
with a real $(1,2,1)$ multiplet and a cubic potential. 
The action is real. Depending on the overall sign (``Euclidean" or ``Lorentzian") of the deformation, 
a Bender-Boettcher pseudo-hermitian hamiltonian is encountered when solving the equations of motions of the auxiliary field. A possible connection of our framework with the Drinfeld twist deformation of supersymmetry is pointed out.
\end{abstract}
\vfill 
\rightline{CBPF-NF-009/07}

\newpage
\section{Introduction}
In this paper we investigate the properties of Lie supergroups whose odd-parameters are Clifford-valued (instead of being Grassmann numbers). 
The case of the Supersymmetric Quantum Mechanics (s.t. the Lie superalgebra is given by the one-dimensional $N$-extended supersymmetry algebra)
is explicitly discussed. The extension of the approach to odd-Clifford supergroups based on higher-dimensional super-Poincar\'e superalgebras
is immediate.\par
We produce an extension of the Berezin calculus which takes into account the Clifford property of the odd variables.
The supersymmetric fermionic covariant derivatives are derived with standard methods. It is of particular interest the case of the
$N=2$ one-dimensional supersymmetry. The $Cl(2,0)$ and $Cl(1,1)$ (``Euclidean" and respectively ``Lorentzian") Clifford generalizations
of the ordinary $N=2$ one-dimensional Grassmann supersymmetry imply fermionic covariant derivatives which do not satisfy the
graded Leibniz property. In our framework this fact proves to be crucial to produce genuine deformations of the ordinary $N=2$ Supersymmetric Quantum Mechanical models. Moreoveor, depending on the type of Clifford deformation and the chosen values of the coupling constants,
we naturally induce Bender-Boettcher ${\cal PT}$-symmetric pseudo-hermitian hamiltonians \cite{bb} from real $N=2$ supersymmetric actions. \par
Supergroups with odd Clifford parameters turn to be a very natural framework to describe Non-anticommutative supersymmetric theories. We recall that non-anticom-mutative supersymmetry has received considerable attention in the last few years.
A number of papers \cite{{svn},{fl},{flm},{kpt},{ov},{ov2},{sei}} have explored the implications of introducing non-anticommutative spinorial coordinates, either as a mathematical possibility
or in the string context (see \cite{{gpr},{is}} for recent reviews). The  work of \cite{sei}, introducing a non-anticommutative supersymmetry in a 4-dimensional
Euclidean superspace, has been particular influential. The construction of nonanticommutative supersymmetric theories in lower dimensions
(two, see \cite{{ck},{cha},{hkks},{hkks2},{agvm}}) or three (see \cite{fgnps}) has later been investigated. In \cite{as} it was pointed out that the one-dimensional framework of
the $N=2$ non-anticommutative supersymmetric quantum mechanics could be important for understanding several mathematical properties of non-anticommutative
supersymmetry, as well as exploring physical applications (e.g. to condensed matter physics).\par
The majority of the recent works on non-anticommutative supersymmetry has been inspired by the non-commutative deformation of the ordinary bosonic
theories. Due to this reason, the most employed tool is a deformed Moyal star product applied to fermionic (anticommuting) variables.
In \cite{{hh},{hhs}} for instance, the deformed quantization program of \cite{bffls} is taken as an inspiration to construct Clifford algebras from Moyal star-products of
Grassmann generators.\par
In this paper we are advocating a somehow complementary viewpoint. We start, from the very beginning, with a Clifford algebra, whose properties are known. The squares of the odd-generators have a mass-dimension $mass^{-1}$; they are therefore naturally associated to a Clifford-deformation mass scale $M$. By letting $M\rightarrow\infty$ we are able to recover, in the limit, the Grassmann case.
\par
It is worth recalling that several different prescriptions have been given in the literature to introduce non-anticommutative deformations of ordinary
supersymmetry. In most of the cases, the supersymmetry algebra itself is deformed (see e.g. \cite{sei}). On the other hand, as it was already pointed
out in \cite{sei}, the supersymmetry algebra can be restored at the price of introducing fermionic covariant derivatives which do not
satisfy the graded Leibniz property. In \cite{sei} and several other papers following it, the investigation is restricted to graded Leibniz derivatives
which obey the graded Leibniz property. Two main motivations for that are given. The first one concerns chiral (antichiral) superfields; the breaking of Leibniz implies
that the product of chiral superfields is no longer chiral. The second motivation concerns the impossibility of integrating by parts. These two
motivations can be easily overcome, at least in selected cases. There are interesting theories which do not require the presence of chiral or
antichiral superfields. The $N=2$ supersymmetric quantum mechanics for the {\em real} $(1,2,1)$ superfield (with one auxiliary field),
discussed in Section {\bf 6},  is an example.
Moreover, for this kind of theory, the supersymmetric potentials are manifestly supersymmetric invariant, because the supersymmetry generators,
applied to the integrand, produce a total time-derivative applied to the only term surviving the integration over odd-Clifford variables. \par
We postpone to the Conclusions further discussions of several features of our approach. These features include the connection with pseudohermitian
hamiltonians, the connection between the breaking of the graded Leibniz property and the twisted deformation of supersymmetry, the extension of our construction to higher dimensions.\par
The scheme of this paper is as follows. In Section {\bf 2} 
supergroups with odd Clifford parameters are introduced. The supergroups associated to the superalgebras of
the one-dimensional $N$-extended supersymmetry are explicitly discussed. In Section {\bf 3} a Berezin-like calculus
is presented for odd variables which are no longer anticommuting (Grassmann). In Section {\bf 4} we derive the fermionic
covariant derivatives for superspaces with odd Clifford variables. The superfield formalism for
such superspaces is introduced in Section {\bf 5}. The one-dimensional $N$-extended supersymmetry
is investigated and the conditions under which the fermionic covariant derivatives obey the graded Leibniz rule are
expressed. The properties of the ``Euclidean" and ``Lorentzian" Clifford deformations of the $N=2$ one-dimensional supersymmetry are
discussed.  In Section {\bf 6} the odd-Clifford approach is employed to introduce the Non-Anticommutative Supersymmetric
Quantum Mechanics. A detailed analysis of the $N=2$ SQM in terms of a real $(1,2,1)$ superfield with a trilinear superpotential
is made. The auxiliary field satisfies (for Euclidean and Lorentzian deformations) an algebraic equation of motion.
In the purely bosonic limit the theory is described by a trilinear potential. Depending on the type of deformation and
the value of the coupling constant, the effective hamiltonian can be reduced to a Bender-Boettcher ${\cal P T}$-symmetric 
pseudo-hermitian hamiltonian. In the Conclusions we discuss several features of our construction, such as the connection with pseudo-hermitian hamiltonians, the possibility of interpreting the breaking of the graded Leibniz rule of the fermionic
covariant derivatives as a non-trivial coproduct within the Drinfeld twist deformation of the supersymmetry, etc. 
The necessary modifications to accommodate within this framework higher-dimensional supersymmetric theories are mentioned.

\section{Supergroups with odd Clifford parameters: the one-dimensional $N=1,2$ supersymmetry}

Lie superalgebras are ${\bf Z}_2$-graded algebras whose generators, split into even and odd sectors,
satisfy (anti)-commutation relations (see \cite{{kac1},{kac2},{fss}} for a precise definition).
Examples of Lie superalgebras include the algebra of the one-dimensional $N$-extended supersymmetry
discussed below, the super-Poincar\'e algebra, the simple Lie superalgebras (\cite{fss}).
For ordinary Lie groups the elements connected with the identity are obtained through
``exponentiation" of the Lie algebra generators. Similarly, the elements connected to the identity of the Lie supergroups
are obtained by exponentiating the Lie superalgebra generators. In the standard construction (\cite{{kac1},{kac2},{bt}}),
Lie supergroups are locally expressed in terms of bosonic parameters associated to the even generators of the Lie superalgebra, while fermionic, Grassmann-number parameters are associated to the odd generators (being Grassmann, in particular, their
square is assumed to vanish).\par
It is tempting to understand the non-anticommutative formulation of the supersymmetry by relaxing the Grassmann condition
for the odd parameters. In the examples here discussed we assume them to satisfy a more general class of algebras.
Let's take $N$ odd parameters $\theta_i$ ($i=1,\ldots, N$); we can assume their anticommutators (once conveniently normalized)
being expressed through
\begin{eqnarray}\label{thetas}
\theta_i\theta_j+\theta_j\theta_i &=& 2\eta_{ij},
\end{eqnarray}
where $\eta_{ij}$ is a diagonal matrix with $p$ elements $+1$, $q$ elements $-1$ and $r$ zero elements in the diagonal
(therefore $N=p+q+r$). The subsector of $r$ $\theta_i$'s with vanishing square is still Grassmann. For $r=0$,
the equation (\ref{thetas}) reduces to the basic relation of the generators of the $Cl(p,q)$ Clifford algebra.\par
We can refer to the procedure of replacing Grassmann variables with the (\ref{thetas}) relation as ``Cliffordization"
of the supergroup elements.  The corresponding supersymmetry will be denoted as ``$Cl(p,q,r)$-type". The ordinary supersymmetry
is recovered for $p=q=0$ (therefore, it is of ``$Cl(0,0,N)$-type"). The Cliffordization is here proposed as a framework to understand the features of the ``non-anticommutative supersymmetry".
In a different context than the one here discussed, supertransforamtions not depending on Grassmann parameters were also investigated in \cite{int}.
\par
In this paper we are mostly concerned with the example of the one-dimensional 
$N$-extended supersymmetry algebra underlying the Supersymmetric Quantum Mechanics. It is explicitly given in terms of a single 
even generator $H$ (the hamiltonian, in physical applications) and $N$ odd generators $Q_i$ ($i=1,\ldots , N$), satisfying
the (anti)commutation relations
\begin{eqnarray}\label{Nsusy}
\{ Q_i,Q_j\}&=&2\delta_{ij} H,\nonumber\\
\relax [H, Q_i]&=& 0.
\end{eqnarray}
Let us discuss first the $N=1$ case ($Q^2=H$). The bosonic parameter associated with $H$ is the ``time" $t$, while, for later convenience, we denote as $\theta_\lambda$ the odd-parameter associated with $Q$. We notice that $\theta_\lambda$ can be expressed as
$\theta_\lambda= \lambda\theta$, where $\lambda$ is a real number and $\theta$ is a given odd-parameter of reference. \par
For both even and odd variables $a$, $b$, the conjugation ``$\ast$" is defined (see \cite{{ber},{ber2}}) satisfying
\begin{eqnarray}
(ab)^\ast &=& b^\ast a^\ast,
\nonumber\\
(a^\ast)^\ast &=& a.
\end{eqnarray}
Accordingly, an element $g$ of the unitary $N=1$ supergroup is expressed as 
\begin{eqnarray}\label{unitarygroup}
g&=& e^{-iHt}e^{\lambda\theta Q}.
\end{eqnarray}
We assumed the reality condition 
\begin{eqnarray}
\theta^\ast&=&\theta.
\end{eqnarray}
In mass-dimension, we have
\begin{eqnarray}
&\begin{array}{cc}{}
\relax [H] =1, & [t]=-1,\\
\relax [Q] =\frac{1}{2}, & [\theta_\lambda] =-\frac{1}{2}.
\end{array}
&\end{eqnarray}
Since $\theta^2$ must have the correct mass-dimension, it should be expressed in terms of some 
positive mass scale $M$. We can distinguish three cases, up to a normalization factor. We can set
\begin{eqnarray}
\theta^2 &=& \frac{\epsilon}{M},
\end{eqnarray}
with $\epsilon =0$ (the Grassmann case), $\epsilon=+1$ or $\epsilon=-1$.
In terms of $\theta_\lambda$ we have that each $\theta_\lambda$ satisfy, in the three respective cases, the conditions ${\theta_\lambda}^2=0$,  ${\theta_\lambda}^2>0$ or ${\theta_\lambda}^2<0$.\par
As a result we obtain three $N=1$ supergroups associated to the $N=1$ superalgebra (\ref{Nsusy}). By
setting
\begin{eqnarray}
X &=& \sqrt{\frac{\lambda^2 H}{M}}
\end{eqnarray}
we obtain, for $\epsilon =0,1,-1$:
\\
{\em i}) $\epsilon =0$, the {\em rational} case,
\begin{eqnarray}
g &=& e^{-iHt}(1+\lambda\theta Q),
\end{eqnarray}
{\em ii}) $\epsilon =1$, the {\em trigonometric} case,
\begin{eqnarray}
g &=& e^{-iHt}(\cos X+I\sin X),
\end{eqnarray}
\quad\quad where $
I=\theta Q\sqrt{\frac{M}{H}}$ and $I^2=-1$;\\
{\em iii}) $\epsilon =-1$, the {\em hyperbolic} case,
\begin{eqnarray}
g &=& e^{-iHt}(\cosh X+J\sinh  X),
\end{eqnarray}
\quad\quad where $
J=\theta Q\sqrt{\frac{M}{H}}$ and $ J^2=1$.

Notice that the ordinary Grassmann case can be recovered from both Clifford cases in the special limit
$M\rightarrow\infty$.\par
The extension of the above procedure for arbitrary $N$ is straightforward. In the following we are mostly interested to the $N=2$ case. It is characterized by two odd parameters of reference, $\theta_1$,
$\theta_2$(${\theta_i}^\ast=\theta_i$), satisfying ${\theta_i}^2=\frac{\epsilon_i}{M}$, where both $\epsilon_1$, $\epsilon_2$ can assume the
three values $0$, $+1$ and $-1$. As it will appear in the following, most of the interesting properties of the
Cliffordized $N=2$ supersymmetry are expressed in terms of the product
\begin{eqnarray}\label{epsilon}
\epsilon &=&\epsilon_1\epsilon_2.
\end{eqnarray} 
The three cases for $\epsilon$ correspond to\\
{\em i}) $\epsilon=0$, where at least one of the two $\theta$'s is Grassmann,\\
{\em ii}) $\epsilon =+1$, the ``Euclidean" version of the $N=2$ Clifford Supersymmetry (realized by either
$Cl(2,0)$ or $Cl(0,2)$),\\
{\em iii}) $\epsilon=-1$, the ``Lorentzian" version of the $N=2$ Clifford Supersymmetry 
(obtained for $Cl(1,1)$).

\section{A Berezin-like calculus for odd-Clifford variables}
The Berezin calculus sets the rules for the derivation and the integration of odd Grassmann variables \cite{ber}.
For our purposes we need to introduce a calculus which substitutes the Berezin calculus in the case of
odd variables of Clifford type.\par
For a single Grassmann variable $\theta$ the Berezin calculus states that the derivative $\partial_\theta$
is normalized s.t. $\partial_\theta \theta = 1$, while giving vanishing results otherwise. 
The Berezin integration $\int d\theta$ coincides with the Berezin derivation ($\int d\theta=\partial_\theta$). The extension of the Berezin calculus to an arbitrary number of Grassmann variables
is straightforward.
Notice that, if $\theta$ has mass-dimension $[\theta]=-\frac{1}{2}$ as in the previous Section, then
$\partial_\theta$ has mass-dimension $[\partial_\theta ]=\frac{1}{2}$.\par
We establish now the rules for an analogous calculus in the case of an odd $\theta$ s.t.
$\theta^2 = \frac{\epsilon}{M}\neq 0$.
We introduce an odd derivation $\partial_\theta$ for the Clifford $\theta$ by assuming that
it has the same mass-dimension as the Berezin derivation and coincides with it in the $M\rightarrow\infty$ limit. Therefore
\begin{eqnarray}
\partial_\theta\theta&=& 1,\nonumber\\
\partial_\theta 1&=& 0.
\end{eqnarray}
The application of $\partial_\theta$ to the powers $\theta^n$, for integral values $n>1$, is determined under
the assumption that $\partial_\theta$ satisfies a graded Leibniz rule. Let $f_1$, $f_2$ be two functions of
grading $deg(f_1)$, $deg(f_2)$ respectively ($deg(f)=0$ for a bosonic function $f$, while $deg(f)=1$ for a fermionic function); we assume 
\begin{eqnarray}
\partial_\theta (f_1f_2) &=& (\partial_\theta f_1)f_2 + (-1)^{-deg(f_1)}f_1(\partial_\theta f_2).
\end{eqnarray}
As a consequence, the application of $\partial_\theta$ to even and odd powers of $\theta$ is respectively given by
\begin{eqnarray}
\partial_\theta \theta^{2k}&=& 0,\nonumber\\
\partial_\theta \theta^{2k+1}&=& \theta^{2k}.
\end{eqnarray}
By requiring the integral of a total derivative to be vanishing we can unambiguously fix
\begin{eqnarray}\label{thetaint1}
\int d\theta ~\theta^{2k}&=& 0.
\end{eqnarray}
The rule for the integration over odd powers of $\theta$ can be set by requiring, as in the Berezin case,
that the integration coincides with the derivation $\partial_\theta$. Therefore
\begin{eqnarray}\label{thetaint2}
\int d\theta ~\theta^{2k+1} &=& \theta^{2k}.
\end{eqnarray}
There is an extra reason motivating (\ref{thetaint2}) as the correct
prescription for the integration. The even powers of $\theta$ are bosonic (even) elements which can be expressed in terms of the
mass $M$, which is expected to play a physical role. We recall that $\theta^{2k} = \frac{\epsilon^k}{M^k}$.
The (\ref{thetaint2}) prescription allows to perform the substitution $\theta^{2k+1}=\frac{\epsilon^k}{M^k}\theta $ 
and treat $\frac{\epsilon^k}{M^k}$ as an ordinary bosonic parameter, unaffected by the odd integration.\par
With the above (\ref{thetaint1}) and (\ref{thetaint2}) prescriptions, the derivation and integration over an odd Clifford variable are {\em formally} similar to the Berezin counterparts. All $\theta$-valued fields can be regarded as at most linear in
$\theta$, so that the standard Berezin rules for derivation and integration apply. The difference w.r.t. the usual
Grassmann case lies in the product of $\theta$-valued fields, since extra contributions arise from the non-vanishing of $\theta^2$. A $d$-dimensional bosonic $\theta$-valued field $\Phi$ can be expressed as
\begin{eqnarray}
\Phi (\theta) &=& \phi + i\psi \theta.
\end{eqnarray} 
In the usual Grassmann case it corresponds to a bosonic component field $\phi$ of mass dimension $[\phi]= d$,
plus its fermionic counterpart $\psi$ of mass-dimension $[\psi]= d+\frac{1}{2}$. In the Clifford case, for
$\epsilon=\pm 1$, the bosonic field $\phi$ is Taylor-expanded in powers of $\frac{1}{M}$:
\begin{eqnarray}
\phi &=& \sum_{n=0}^{+\infty} \frac{\phi_n} {M^n}.
\end{eqnarray}
Its $\phi_n$ subcomponents have mass-dimension $[\phi_n]=d+n$. The Grassmann case is recovered by the $\phi_0$
subcomponent which survives when taking the $M\rightarrow \infty$ limit. The fermionic field $\psi$ is similarly treated.\par
The extension of the calculus to $N=p+q+r$ odd variables of $Cl(p,q,r)$-type is immediate. In the following
we will work with $N=2$. The two derivatives $\partial_{\theta_1}$, $\partial_{\theta_2}$ satisfy
$\partial_{\theta_i}1=0$, $\partial_{\theta_i}\theta_j =\delta_{ij}$. The double integration
$\int \int d\theta_1 d\theta_2$ is only non-vanishing when applied to the product of $\theta_1$, $\theta_2$
(the even powers of $\theta_i$'s are assumed to be replaced in the integrand by the powers in $\frac{1}{M}$,
${\theta_i}^{2k}=\frac{\epsilon_i}{M^k}$, as explained above):
\begin{eqnarray}\label{N2int}
\int\int d\theta_1 d\theta_2 ~ ~\theta_2\theta_1 &=& 1.
\end{eqnarray}

\section{Fermionic covariant derivatives for Clifford-valued superfields}

In \cite{ss} (see also \cite{wb}) the construction of supersymmetric fermionic covariant derivatives for 
Grassmann variables was discussed. A similar procedure is now adopted to derive supersymmetric fermionic 
covariant derivatives in the case of odd-Clifford variables. For simplicity let's start discussing the $N=1$ supersymmetry
algebra (\ref{Nsusy}). Its supergroup element is given by $g$, introduced in (\ref{unitarygroup}). It is convenient, for the moment,
to keep explicit the dependence on the $\lambda$ parameter. The left (right) action of the supersymmetry generator $Q$ on $g$
($Qg$ and, respectively, $gQ$) induces the operator $Q_L$ ($Q_R$), determined in terms of $t,\theta,\lambda$ and their derivatives, s.t. 
\begin{eqnarray}\label{leftrightaction}
Qg &=& Q_L g,\nonumber\\
gQ &=& Q_R g,
\end{eqnarray}
where
\begin{eqnarray}\label{algebra}
\{Q_L,Q_L\} &=& -H,\nonumber\\
\{Q_L,Q_R\}&=&0,\nonumber\\
\{Q_R,Q_R\}&=& H.
\end{eqnarray}
$Q_L$ is the covariant fermionic derivative, also denoted as ``$D$", while $Q_R\equiv Q$.\par 
In the Grassmann ($\epsilon=0$) case,
$Q_L$, $Q_R$ are explicitly given by
\begin{eqnarray}
Q_L &=& \frac{1}{\lambda}{\partial_\theta} - i\lambda\theta\frac{\partial}{\partial t},\nonumber\\
Q_R &=& \frac{1}{\lambda}{\partial_\theta} + i\lambda\theta\frac{\partial}{\partial t}.
\end{eqnarray}  
The (\ref{algebra}) algebra is consistently reproduced by setting $\lambda=1$, allowing in the Grassmann case to
deal with an $N=1$ superspace depending only on the time parameter $t$ and the Grassmann variable $\theta$. The hamiltonian
$H$ is expressed as $H=i\frac{\partial}{\partial t}$. We have
\begin{eqnarray}
D &=& {\partial_\theta} - i\theta\frac{\partial}{\partial t},\nonumber\\
Q &=& {\partial_\theta} + i\theta\frac{\partial}{\partial t}.
\end{eqnarray} 
In the odd-Clifford case (for $\epsilon=\pm 1$), a solution to the (\ref{leftrightaction}) equations is provided by
\begin{eqnarray}
Q_L &=& \partial_\theta{\partial_\lambda}-i{\theta}\partial_t \int d\lambda + i{\frac{\epsilon}{M}}
\partial_\theta\partial_t\int d\lambda , \nonumber\\
Q_R &=& \partial_\theta{\partial_\lambda}+i{\theta}\partial_t \int d\lambda - i{\frac{\epsilon}{M}}
\partial_\theta\partial_t\int d\lambda.
\end{eqnarray}
When $Q_L$, $Q_R$ are constrained to be applied to superfields whose dependence on $\lambda$ is given by $\exp(\lambda)$, then both the $\lambda$-derivation $\partial_\lambda$ and the
$\lambda$-integration $\int d\lambda$ act as identity. The (\ref{algebra}) algebra is reproduced by $D\equiv Q_L$, $Q\equiv Q_R$ (with dropped $\lambda$-dependence), given by 
\begin{eqnarray}\label{DQ}
D &=& \partial_\theta-i{\theta}\partial_t+ i{\frac{\epsilon}{M}}
\partial_\theta\partial_t ,\nonumber\\
Q &=& \partial_\theta+i{\theta}\partial_t  - i{\frac{\epsilon}{M}}
\partial_\theta\partial_t.
\end{eqnarray}
They are, respectively, the fermionic covariant derivative and the supersymmetry generator expressed in terms of an 
$N=1$ superspace parametrized by the time $t$ and an odd Clifford variable $\theta$ s.t. $\theta^2=\frac{\epsilon}{M}$.  
The extra terms (proportional to $\frac{1}{M}$) appearing on the r.h.s. of the odd Clifford case w.r.t. the Grassmann case
have a purpose. They compensate for the non-vanishing $\theta^2$ to provide the correct supersymmetry transformations
for the component fields of an $N=1$ superfield. Set a bosonic superfield $\Phi=\phi+i\psi\theta$. Its supersymmetry transformation
$\delta_\varepsilon\Phi=\varepsilon Q \Phi$ gives, for its component fields in both Grassmann and odd-Clifford cases,
\begin{eqnarray}
\delta_\varepsilon\phi&=& - i\varepsilon \psi,\nonumber\\
\delta_\varepsilon\psi&=& \varepsilon\partial_t\phi.
\end{eqnarray}
The generalization of the above construction to the $N=2$ case with two odd Clifford variables $\theta_j$ (${\theta_j}^2=
\frac{\epsilon_j}{M}$, $j=1,2$) is immediate. The two fermionic covariant derivatives $D_j$ and the two supersymmetry generators $Q_j$ are respectively given by
\begin{eqnarray}\label{N2derivatives}
D_j&=& \partial_{\theta_j} -i\theta_j\partial_t+i\frac{\epsilon_j}{M}\partial_{\theta_j}\partial_t,\nonumber\\
Q_j&=& \partial_{\theta_j} +i\theta_j\partial_t-i\frac{\epsilon_j}{M}\partial_{\theta_j}\partial_t.
\end{eqnarray}
They satisfy the algebra
\begin{eqnarray}\label{algebraN2}
\{D_i,D_j\} &=& -\delta_{ij}H,\nonumber\\
\{D_i,Q_j\}&=&0,\nonumber\\
\{Q_i,Q_j\}&=& \delta_{ij}H.
\end{eqnarray}

\section{The $1D$ $N=1$ and $N=2$ superfields in the odd Clifford formalism}

Let's denote with $A_k$ a set of $N=1$ superfields expanded in the odd Clifford variable $\theta$. 
The grading $deg(A_k)$ specifies the bosonic ($deg(A_k)=0$) or fermionic ($deg(A_k)=1$) character of $A_k$.\par
In the odd Clifford case the ordinary superfield multiplication must be replaced by the (anti)symmetrized
$\ast$-multiplication defined as follows
\begin{eqnarray}\label{starmultiplication}
A_1\ast A_2 &=& \frac{1}{2}(A_1A_2+ (-1)^{deg(A_1)deg(A_2)} A_2A_1).
\end{eqnarray}
There are several reasons motivating (\ref{starmultiplication}) as the correct prescription. We notice at first
that the $\ast$-multiplication
preserves the reality condition. If $A_1,A_2$ are real, then $A_1\ast A_2$ is real for $deg(A_1)deg(A_2)=0$,
imaginary for $deg(A_1)deg(A_2)=1$.\par The $\ast$-multiplication induces the $\ast$-(anti)commutation relations
defined through
\begin{eqnarray}\label{starmult}
\relax [A_1,A_2\}_\ast &=& A_1\ast A_2 -(-1)^{deg(A_1)deg(A_2)} A_2\ast A_1.
\nonumber
\end{eqnarray}
$\relax [A_1,A_2\}_\ast $ is always vanishing, guaranteeing that the superfields $\ast$-(anti)commute.
\par
The $N=1$ covariant derivative $D$ (\ref{DQ}) satisfies a graded Leibniz property w.r.t. the $\ast$-multiplication.
Indeed
\begin{eqnarray}
D(A_1\ast A_2)&=& (D A_1)\ast A_2 +(-1)^{deg(A_1)} A_1\ast (DA_2).
\end{eqnarray}
The $N=2$ supersymmetry requires the introduction of two, $\theta_1$, $\theta_2$, odd variables. It admits two
irreducible representations \cite{pt}, the real (also denoted as $(1,2,1)$, with one auxiliary field) and the
$(2,2)$ chiral representation.\par 
A real bosonic $N=2$ superfield $\Phi$ is given by
\begin{eqnarray}\label{realN2}
\Phi &=& \phi +i\psi_1\theta_1+i\psi_2\theta_2+ i f \theta_1\theta_2,
\end{eqnarray}
with real bosonic, $\phi$ and $f$, component fields of mass-dimension $d$ and $d+1$, respectively ($f$ is the auxiliary
field).
The real component fermionic fields $\psi_1$ and $\psi_2$ have mass-dimension $d+\frac{1}{2}$.
\par
The chiral $(2,2)$ representation is realized in terms of constrained {\em complex} superfields. Let $\Upsilon$ denote a complex
superfield. In terms of the $N=2$ covariant derivatives ${D}$, $\overline{D}$, given by 
\begin{eqnarray}
{D} &=& D_1-i D_2,\nonumber\\
\overline{D}&=& D_1+iD_2,
\end{eqnarray}
where $D_1$,$D_2$ have been introduced in (\ref{N2derivatives}), the chirality condition for $\Upsilon$ reads as
\begin{eqnarray}
\overline{D} \Upsilon &=& 0
\end{eqnarray}
(the antichirality condition is obtained by replacing $\overline{D}$ with ${D}$).\par
For $\epsilon_1=\epsilon_2=\rho=\pm 1$ (namely, the Euclidean $N=2$ Clifford supersymmetry described at the end of Section {\bf 2}) the bosonic chiral superfield $\Upsilon$ is
expressed in terms of its complex component fields $\varphi$, $\xi$ as
\begin{eqnarray}
\Upsilon&=& \varphi+i\frac{\rho}{M}\dot{\varphi} +\xi\theta-\frac{i}{2}\dot{\varphi}\theta{\overline\theta}
\end{eqnarray}
(here $\theta= \theta_1+i\theta_2$, ${\overline\theta}=\theta_1-i\theta_2$, while the dot denotes, as usual, the time-derivative). \par
The (anti)symmetrized $\ast$-multiplication is introduced for $N=2$ superfields as in $N=1$. However, unlike the $N=1$ case, the $N=2$ covariant derivatives $D_1$, $D_2$ do not satisfy a graded Leibniz property for $\epsilon=\epsilon_1\epsilon_2\neq 0$ (see (\ref{epsilon})). In order to preserve a graded Leibniz property, for $\epsilon\neq 0$, the $\ast$-multiplication must be modified with an extra
term proportional to $\frac{1}{M^2}$. Given two $N=2$ superfields $A$, $B$, the $\hat{\ast}$-multiplication,
defined as
\begin{eqnarray}\label{hatast}
A\hat{\ast}B&=&A\ast B +\frac{\epsilon}{M^2}\partial_{\theta_1}\partial_{\theta_2}A \cdot\partial_{\theta_1}\partial_{\theta_2}B
\end{eqnarray}
is such that it preserves the graded Leibniz property for $D_i$, $i=1,2$,
\begin{eqnarray}
D_i(A\hat{\ast}B)&=&(D_iA)\hat{\ast} B +(-1)^{-deg A} A{\hat{\ast}}D_iB.
\end{eqnarray}

Alternatively, the breaking of the graded Leibniz rule for the $\ast$-multiplication can be expressed as a non-vanishing
$\Delta_i(A,B)$, where
\begin{eqnarray}
\Delta_i(A,B)&=& D_i(A\ast B) -(D_iA\ast B +(-1)^{-deg(A)}A\ast D_iB).
\end{eqnarray}

For bosonic superfields $A$, $B$ s.t. $A=\phi_A+i\psi_{1A}\theta_1+i\psi_{2A}\theta_2+ i f_A \theta_1\theta_2$ and\par
$B=\phi_B+i\psi_{1B}\theta_1+i\psi_{2B}\theta_2+ i f_B \theta_1\theta_2$, $\Delta_1(A,B)$ is, e.g.,
given by
\begin{eqnarray}\label{extraterm}
\Delta_1(A,B)&=& -i\frac{\epsilon}{M^2} (f_B\dot{\psi}_{2A}+f_A\dot{\psi}_{2B} +(f_A{\dot{f}_B}+\dot{f}_Af_B)\theta_1). 
\end{eqnarray}

\section{The Supersymmetric Quantum Mechanics in the Clifford approach}

The constant supersymmetric kinetic term of the $N=1$ superfield (expanded in the odd Clifford variable $\theta$) $\Phi=\phi+i\psi\theta$ ($\theta^2=\frac{\epsilon}{M}$) of mass-dimension $d=0$ is given by
the $N=1$ action
\begin{eqnarray}
S_{N=1}&=& \frac{i}{2m}\int dt \int d\theta (\dot \Phi\ast D\Phi).
\end{eqnarray}
The $N=1$ derivative $D$ is given in (\ref{DQ}), while $\int d\theta$ is the odd Clifford integration specified by (\ref{thetaint1}) and (\ref{thetaint2}).
In component fields the kinetic action reads as
\begin{eqnarray}
S_{N=1}&=& \frac{1}{2m}\int dt ({\dot\phi}^2-i{\dot\psi}\psi)
\end{eqnarray}
and coincides with the $N=1$ constant kinetic action in the Grassmann case. \par
Let us discuss now the $N=2$ Supersymmetric Quantum Mechanics for the real $(1,2,1)$ superfield
$\Phi=\phi+i\psi_1\theta_1+i\psi_2\theta_2+if\theta_1\theta_2$ introduced in (\ref{realN2}).
The two odd Clifford variables $\theta_i$ satisfy ${\theta_i}^2=\frac{\epsilon_i}{M}$. The parameter $\epsilon=\epsilon_1\epsilon_2$
has been introduced in (\ref{epsilon}). The $\int d\theta_1d\theta_2$ integration is defined in (\ref{N2int}). The covariant derivatives
$D_1$, $D_2$ are given in (\ref{N2derivatives}). Concerning the superfields multiplication two options are equally admissible for $\epsilon\neq 0$.
Either superfields are multiplied w.r.t. the (anti)symmetrized ${\ast}$-multiplication (formula (\ref{starmult}) applied to $N=2$ superfields), 
or w.r.t. the modified ${\hat{\ast}}$-multiplication (\ref{hatast}) which guarantees the graded Leibniz rule for $D_i$'s. In both cases
the action, whose integrand is written in terms of superfields and covariant derivatives, is manifestly $N=2$ supersymmetric.
\par
As it happens for deformed Moyal products, due to the property of the $\int d\theta_1d\theta_2 $ integral, bilinear combinations with
${\ast}$-multiplication or ${\hat{\ast}}$-multiplication produce the same results as in the ordinary Grassmann case (${\theta_1}^2={\theta_2}^2=0$).
The effects of the odd Clifford variables can only be detected for $k$-linear products with $k\geq 3$ (trilinear terms and beyond).\par
The $N=2$ free kinetic action of the real superfield $\Phi$ can be written as
\begin{eqnarray}\label{kinN2}
S_{N=2,kin.} &=& \frac{1}{2m} \int dt\int d\theta_1d\theta_2 (D_1\Phi\ast D_2\Phi).
\end{eqnarray}
It reads, in component fields,
\begin{eqnarray}
S_{N=2,kin.} &=&\frac{1}{2m} \int dt({\dot\phi}^2+f^2-i{\dot\psi}_1\psi_1-i{\dot{\psi}}_2\psi_2).
\end{eqnarray} 
The general $N=2$ action is
\begin{eqnarray}
S_{N=2}&=& S_{N=2,kin.}+S_{N=2,pot.},
\end{eqnarray}
where $S_{N=2,pot.}$ is the potential term. \par
For the $N=2$ harmonic oscillator the potential term is quadratic in $\Phi$,
\begin{eqnarray}
S_{N=2, pot.} &=& i\frac{\omega}{2} \int dt\int d\theta_1 d\theta_2 (\Phi\ast\Phi),
\end{eqnarray}
with $\omega$ an adimensional constant.\par
It is required at least a trilinear potential to spot the difference between the Grassmann and the odd Clifford realization
of the $N=2$ supersymmetry.
The most general trilinear potential can be written either as
\begin{eqnarray}\label{astpot}
S_{N=2, pot.} &=& i\int dt \int d\theta_1 d\theta_2 (c_1\Phi\ast\Phi\ast\Phi+ c_2\Phi\ast\Phi+c_3\Phi),
\end{eqnarray}
or
\begin{eqnarray}\label{hatastpot}
{\hat S}_{N=2, pot.} &=& i\int dt \int d\theta_1 d\theta_2 (c_1\Phi {\hat{\ast}}\Phi {\hat{\ast}} \Phi+ c_2\Phi {\hat\ast} \Phi+c_3\Phi).
\end{eqnarray}
The two potentials coincide for $\epsilon=0$.
\par
In (\ref{astpot}) and (\ref{hatastpot}) the coefficients $c_i$'s are real and the $i$ normalizing factor is introduced to ensure
the reality of the $N=2$ potential.\par
Without loss of generality the $c_2$ constant can be set equal to zero ($c_2=0$) through a shift $\Phi\mapsto \Phi'=\Phi+ c$, for a suitable value
$c$. The constant $c_1$ can be normalized s.t.
\begin{eqnarray}
c_1&=&\frac{1}{6},
\end{eqnarray}
leaving the trilinear potential depending on a single real parameter $\alpha=c_3$.\par
In the Grassmann case ($\epsilon_1=\epsilon_2=0$), the full $N=2$ action with the trilinear potential is explicitly given by
\begin{eqnarray}
S_{N=2, Gr.}&=& \int dt({K} - V),
\end{eqnarray}
where ${K}$ is the kinetic term 
\begin{eqnarray}\label{kinetic}
{K} &=& \frac{1}{2m}({\dot\phi}^2-i{\dot\psi}_1\psi_1-i{\dot\psi}_2\psi_2),
\end{eqnarray}
and $V$ is the potential
\begin{eqnarray}
V&=& -(\frac{f^2}{2m}-i\phi\psi_1\psi_2+\frac{1}{2}\phi^2f+\alpha f).
\end{eqnarray}
This result is reproduced in the $\epsilon=0$ odd Clifford case (\ref{astpot}) and, no matter which is
the value of $\epsilon$,
in the (\ref{hatastpot}) case. For the trilinear potential we are guaranteed that, ensuring the graded Leibniz property for the covariant 
fermionic derivatives, the resulting odd Clifford action coincides in components with the ordinary component fields action for Grassmann supersymmetry.
This equivalence is preserved for more general potentials and more general theories. On the other hand, genuine odd Clifford deformations
of the ordinary Grassmann supersymmetry are recovered when the graded Leibniz property of the fermionic covariant derivatives is broken.
In the trilinear example above it corresponds to the choice of the (\ref{astpot}) $N=2$ potential for either $\epsilon=-1$ (the ``Lorentzian" $N=2$
odd Clifford supersymmetry) or  $\epsilon=1$ (the ``Euclidean" $N=2$ odd Clifford supersymmetry). \par
It is worth stressing the result of our analysis, that the Non-Anticommutative Supersymmetry, within the odd Clifford approach, can be understood arising from the breaking of the graded Leibniz property. In the following subsection we discuss in detail the deformed trilinear potentials obtained
for $\epsilon=\pm 1$ (the (\ref{astpot}) prescription is understood) and spot the differences w.r.t. the $\epsilon=0$ case.\par
Before starting this analysis let us point out, as a side remark, that the fermionic component fields entering the superfields are assumed to be Grassmann.
In principle this assumption can be further relaxed, the fermionic fields can be taken as odd Clifford fields with a non-vanishing square.
However, at least for the classes of actions here discussed, no gain is made since the overall effect is reproduced by a shift in the coupling constants
entering the potential term.

\subsection{The $N=2$ trilinear potential for $\epsilon=0,-1,1$}

The trilinear potential, for $\epsilon=\pm 1$, induces an action whose kinetic term $K$ coincides with eq. (\ref{kinetic}),
while the potential $V$ is given by
\begin{eqnarray}
{V} &=& -\left(\frac{1}{2m}f^2+\frac{1}{2}\phi^2f-i\phi\psi_1\psi_2+\frac{1}{6}\frac{\epsilon}{M^2}f^3+\alpha f\right).
\end{eqnarray}
In the $\epsilon=0$ Grassmann supersymmetry, for generic potentials, the equation of motion of the auxiliary field $f$ is
a linear equation. For $\epsilon\neq 0$, the equation of motion for $f$ is an algebraic equation (a second order equation for
the above example of the trilinear potential), with several branches of solutions. The prescription to correctly pick up
a branch is discussed in the following.\par
For simplicity, and without loss of generality, it is convenient to identify the mass-scale $M$ with the mass-scale $m$ 
entering (\ref{kinN2}). We can therefore set $M=m$. The main features of the potential can be understood by taking its purely bosonic
sector, consistently setting all fermionic fields to zero ($\psi_1=\psi_2=0$). 
In the $\epsilon=0$ Grassmann case, solving the equation of motion for $f$ and inserting back into $V$, we obtain
\begin{eqnarray}\label{grpot}
\frac{V}{m}&=& \frac{1}{8}(\phi^2+2\alpha)^2.
\end{eqnarray}
The corresponding theory admits two invariances: supersymmetry and ${\bf Z}_2$-invariance $\phi\mapsto -\phi$.
We can distinguish three cases according to the value of $\alpha$. We have\\
{\em i}) $\alpha>0$: the ${\bf Z}_2$-invariance is exact, while the supersymmetry is spontaneously broken,\\
{\em ii}) $\alpha=0$: both the ${\bf Z}_2$-invariance and the supersymmetry are exact and, finally, \\
{\em iii}) $\alpha<0$: the supersymmetry is exact, while the ${\bf Z}_2$-invariance is spontaneously broken
(the ``mexican hat"-shape potential).\par
This analysis can be repeated for $\epsilon=\pm 1$. We obtain the following equation of motion for the auxiliary field
$f$:
\begin{eqnarray}
f_{\pm} &=& m(-\epsilon \pm \sqrt{1-\epsilon(2\alpha+\phi^2)}.
\end{eqnarray} 
By specializing to $\epsilon=-1$ (the ``Lorentzian" case) and setting
\begin{eqnarray}
x&=& \sqrt{1+2\alpha+\phi^2},
\end{eqnarray}
we obtain two branches for the potential $V$:
\begin{eqnarray}\label{potential}
\frac{V_\pm}{m} &=& \pm\frac{1}{3}x^3-\frac{1}{2}x^2+\frac{1}{6}. 
\end{eqnarray} 
For $\alpha \geq -\frac{1}{2}$,  $x$ is always real. The branches have to be chosen s.t. $V$ is bounded below.
Therefore
\begin{eqnarray}
\frac{V}{m} &=& \frac{1}{3}|x^3|-\frac{1}{2}x^2+\frac{1}{6}. 
\end{eqnarray} 
Three cases have to be distinguished according to the value $\alpha\geq -\frac{1}{2}$. We have
\\
{\em i}) $\alpha>0$: the ${\bf Z}_2$-invariance $\phi\mapsto-\phi$ is exact, while the supersymmetry is spontaneously broken,\\
{\em ii}) $\alpha=0$: both the ${\bf Z}_2$-invariance and the supersymmetry are exact and,\\
{\em iii}) $-\frac{1}{2}\leq\alpha<0$: the supersymmetry is exact, while the ${\bf Z}_2$ invariance is spontaneously broken
(this case corresponds to a deformed version of the ``mexican hat" potential).\par
In the three cases above, $x$ belongs to the real axis. On the other hand $x$ is constrained to satisfy
$|x|\geq \sqrt{1+2\alpha}$ (the whole real axis is recovered for the special value $\alpha=-\frac{1}{2}$).\par
In the Lorentzian $\epsilon=-1$ case, for $\alpha\geq -\frac{1}{2}$, we obtained real potentials which are deformations of the
``Grassmann" potential (\ref{grpot}). 
On the other hand, the reality condition (for the classical theory, the hermiticity condition is understood for its quantum version)
for the $N=2$ odd Clifford action written in terms of the $N=2$ superfield requires $\alpha$ to be an unconstrained real parameter.
In particular, the values $\alpha<-\frac{1}{2}$ are allowed.  In the Lorentzian case, such values correspond to a potential expressed
in terms of $x$, where now $x$ takes value on the whole real axis and on the part of the imaginary axis constrained to
$|x|\leq \sqrt{-2\alpha-1}$.
 \par
 In the Euclidean $\epsilon=1$ case, the two branches of the potential are still furnished by equation (\ref{potential}). On the other
 hand, the $x$ variable is now expressed in terms of the real field $\phi$ as
\begin{eqnarray}
x &=& \sqrt{1-2\alpha -\phi^2}.
\end{eqnarray}
In the Euclidean odd Clifford supersymmetry the $x$ variable always takes some of its values on (part of) the imaginary axis. We can indeed distinguish three 
separate cases according to the value of the $\alpha$ parameter. We have\\
{\em i}) for $\alpha> \frac{1}{2}$, $x$ takes values on the part of the imaginary axis satisfying the constraint
$|x|\geq \sqrt{2\alpha-1}$;   \\
{\em ii}) for $\alpha=\frac{1}{2}$, $x$ takes value on the whole imaginary axis; \\
{\em iii}) for $\alpha<\frac{1}{2}$, $x$ takes value on the whole imaginary axis {\em and} the part of the real axis
satisfying the constraint $|x|\leq \sqrt{1-2\alpha}$.

\subsection{On $N=2$ Clifford-deformed supersymmetry and \\
${\cal PT}$-hamiltonians}

Let us specialize now our discussion to the Euclidean-deformed $\alpha=\frac{1}{2}$ case. For this special choice of $\alpha$,
we have $x=i\phi$, s.t. the purely bosonic effective action $S$ for $\phi$ is given by
\begin{eqnarray}
S&=&  \int dt \left( \frac{1}{2} {\dot\phi}^2 +\frac{i}{3}\phi^3 -\frac{1}{2}\phi^2-\frac{1}{6}\right)
\end{eqnarray}
(we set $m=1$ for simplicity).\par
This action induces a Bender-Boettcher \cite{{bb},{bbm},{ben}} ${\cal{PT}}$-symmetric hamiltonian
$H$ (with $p=\dot{\phi}$)
\begin{eqnarray}
H&=& \frac{1}{2}p^2 -\frac{i}{3}\phi^3+\frac{1}{2}\phi^2+\frac{1}{6},
\end{eqnarray}
invariant under the coupled transformations (see \cite{bbm})
\begin{eqnarray}
{\cal P}&:& \phi \mapsto -\phi,\quad p\mapsto -p,\nonumber\\
{\cal T} &:& \phi\mapsto\phi,\quad\quad p\mapsto -p,\quad i\mapsto -i.
\end{eqnarray}
It is worth stressing the fact that our original odd-Clifford $N=2$ supersymmetric action for $\phi, \psi_1,\psi_2,f$ 
(no matter which Clifford deformation and which real value of the $\alpha$ parameter are taken) satifies the reality condition. It's only after solving the equation of motion
for the auxiliary field $f$ that the imaginary unit $i$ appears (for the above-discussed cases) in the reduced action. What we succeeded here is
to directly link a ${\cal PT}$-symmetric
hamiltonian with a Non-anticommutative $N=2$ supersymmetric quantum mechanical system.\par
For $\alpha\neq\frac{1}{2}$, in terms of the $x$ variable, we get an action with a non-constant kinetic term. The explicit investigation
of the properties of these actions will be left for the future.\par
We conclude this Section mentioning that the trilinear superpotential has been explicitly
discussed here for its simplicity. The $N=2$ odd-Clifford framework for the real ($1,2,1$) multiplet allows the construction of actions for
a general class of superpotentials, whose properties can be analyzed within the scheme here outlined.

\section{Conclusions}

Some of the issues of the Non-anticommutative supersymmetry based on supergroups with odd-Clifford variables deserve further comments.\par
To our knowledge, the first paper mentioning a connection between Non-anticom-mutative supersymmetry and ${\cal PT}$-symmetric (pseudohermitian) hamiltonians
is \cite{ivsm}. In that work, a model introduced in \cite{as} was investigated in detail. The pseudohermitian property of the hamiltonian
(called ``cryptoreality" in \cite{ivsm}) was discussed in terms of the \cite{{mos},{mos2},{mos3},{mos4}} conjugation transformation ${\widetilde H} =e^RHe^{-R}$
relating the pseudohermitian hamiltonian $H$ to
its self-adjoint ${\widetilde H}$ counterpart. In \cite{ivsm} it was further pointed out that such ``cryptoreal" hamiltonians, with real spectrum and a unitary evolution operator, could define a consistent supersymmetric Non-anticommutative theory in a Minkowski space-time. 
In the \cite{mos} approach, the key issue to the reality of the spectrum of the pseudohermitian hamiltonians is the existence of the conjugation
transformation, rather than the presence of a ${\cal PT}$-symmetry. On the other hand, as we have seen, our $N=2$ odd-Clifford framework to Non-anticommutative supersymmetry provides in a very natural way pseudohermitian hamiltonians. Essentially, the complexity of the hamiltonian
is ``artificially induced" by solving the equation of motion of the auxiliary field. The action, written in terms of the whole set of $N=2$ component fields, is real. It is therefore quite natural to pose the question whether the pseudohermitian property of a generic hamiltonian could be recovered 
from the existence of an underlying extended non-anticommutative supersymmetry. Due to the growing importance of the investigations on
pseudohermitian hamiltonians, this issue deserves a careful investigation. \par
Concerning the violation of the graded Leibniz property for covariant fermionic derivatives based on odd-Clifford supersymmetries, 
some works, discussing related results, should be signaled \cite{{ks},{zup},{ihs}}. In these works the non-anticommutative supersymmetry is formulated as a Drinfeld twist deformation of the Hopf algebra structure of (a given) supersymmetry algebra.  The twist deformation implies, in particular,
a deformed coproduct $\Delta$ for the fermionic generators $Q_i$ of the superalgebra, s.t. 
$\Delta (Q_i) = Q_i\otimes {\bf 1} + {\bf 1} \otimes Q_i +(\ldots)$, where $(\ldots)$ denotes the extra terms arising from the deformation.  In
\cite{{lust},{luk}} it was discussed a physical interpretation of the coproduct in the construction of tensored multiparticle states. Let $g$ be a bosonic Lie algebra generator associated to, let's say, a hamiltonian $H$, the undeformed coproduct $\Delta_0(g)=g\otimes{\bf 1} +{\bf 1}\otimes g$ is interpreted,
e.g., as the addition of energy for a two-particle state ($E_{1+2}=E_1+E_2$).
The results of \cite{{ks},{zup},{ihs}} admit the physical interpretation
that the supersymmetry transformation $\delta_\epsilon$ acting on the product of two (let's say bosonic) superfields $\Phi_1$, $\Phi_2$ is such that,
in the deformed case, $\delta_\epsilon(\Phi_1\Phi_2)\neq (\delta_\epsilon\Phi_1)\Phi_2+\Phi_1(\delta_\epsilon\Phi_2)$. 
These results suggest a possible link between the Clifford-deformation approach here discussed and the supersymmetric version of the Drinfeld twist deformation. Indeed, by adapting the results of \cite{{ks},{ihs}} we can check that,
at least for the $N=2$ Euclidean Clifford  deformation, a twist element ${\cal F}$ can be expressed as
${\cal F} = exp(\alpha Q\otimes {\overline Q})$, with $Q=Q_1+iQ_2$, ${\overline Q}=Q_1-iQ_2$. The nilpotency of $Q, {\overline Q}$ implies that ${\cal F}$  is a finite sum. 
By setting $\alpha=-\frac{\epsilon_1}{M}=-\frac{\epsilon_2}{M}$ we obtain Euclidean-deformed Clifford variables as 
$\star$-product of the $\theta_1$, $\theta_2$ Grassmann variables:
$\theta_i\star\theta_j = m\circ {\cal F}^{-1} (\theta_i\otimes \theta_j)$
($m$ is the ordinary multiplication, see \cite{{ks},{ihs}}). 
On the other hand, to make explicit the connection between the component fields entering the Clifford approach
(which, we remember, are Taylor-expanded in $\frac{1}{M}$-powers) and the component fields entering the 
supersymmetric Drinfeld twist framework, would require to find suitable dressing transformations. One cannot in fact naively express both sets of superfields with the same field components. This would require, e.g. to satisfy
the equation (for real $N=2$ superfields $\Phi_A, \Phi_B$), 
$ Q_1 (\Phi_A\ast\Phi_B) = Q_1( \frac{1}{2} m\circ {\cal F}^{-1} (\Phi_A\otimes \Phi_B) + (A\leftrightarrow B))$  
where the l.h.s is the $Q_1$ supersymmetry transformation in the Clifford approach, while the r.h.s. corresponds to
 the supersymmetry transformation in the twist Drinfeld framework.  It can be easily check that the above equality
 is not satisfied. This result does not rule out, altogether, a possible connection between Clifford-deformation and Drinfeld twist. It simply points out that it could be obtained through suitable dressing trasnformations.
This is a very interesting line of investigation which we are planning to address in forthcoming papers.
\par
In this work we have explicitly discussed supergroups based on odd Clifford variables associated to the one-dimensional $N$-extended superalgebra.
The generalization of the present construction to superPoincar\'e algebras in higher dimensions is straightforward. The odd generators are spinors and
carry spinorial indices. To preserve the Lorentz covariance, in the real case, the odd cordinates $\theta_\alpha$ must satisfy anticommutation relations
such as $\{\theta_\alpha,\theta_\beta\} = C_{\alpha\beta}$, where $C_{\alpha\beta}$ is a constant charge conjugation matrix, symmetric in the $\alpha\leftrightarrow\beta$ exchange. This construction is only possible for space-times, see \cite{{kuto},{dat},{crt}}, admitting a symmetric charge conjugation matrix. Alternatively, for complex, Dirac, odd coordinates, non-vanishing constant anticommutation relations can be imposed in the presence of a
hermitian $A$ matrix discussed in \cite{{kuto},{dat},{crt}}. The Weyl projection, when needed, can also be accommodated in this framework. The extension of the Berezin calculus to these cases follows from the rules of the odd-Clifford calculus here discussed. The fermionic covariant derivatives can be similarly obtained. 
\par  
The present results can find several possible applications. Quite naturally, the $N=2$ odd-Clifford one-dimensional supersymmetry 
can be discussed in the context of the deformation of the non-relativistic $N=2$ supersymmetric integrable systems in $1+1$-dimensions
(such as the $N=2$ KdVs and analogous $N=2$ KP-reduced hierarchies). Furthermore, the analysis of two-dimensional superconformal models is particularly interesting
in order to understand the role played by the mass-scale $M$ entering the Clifford relations for odd variables. It is also quite
natural to extend the present
investigation to four or higher-dimensional supersymmetric field theories. 

{}~
\\{}~
\\
 
{\large{\bf Acknowledgments}}{} ~\\{}~\\
This work received support from CNPq (Edital Universal 19/2004) and FAPERJ. \\
Z. K. acknowledges
FAPEMIG for financial support and CBPF for hospitality. \\
M. R. acknowledges financial support
from CLAF.

\end{document}